%% file: henning_cipanp09.tex
\input{aipcheck}

\documentclass[
    ,final            
    ,numberedheadings
  ]
  {aipproc}

\usepackage{graphicx}
\usepackage{fancyhdr}

\def\BBz{$\beta\beta(0\nu)$}

\def\BBt{$\beta\beta(2\nu)$}

\def\MJ{{\sc Majorana}}             
\def\be{\begin{equation}}
\def\ee{\end{equation}}

\def\gess{$^{76}\mathrm{Ge}$}

\def\MJ{{\sc Majorana}}             
\def\be{\begin{equation}}
\def\ee{\end{equation}}

\def\gess{$^{76}\mathrm{Ge}$}

\layoutstyle{6x9}

\renewcommand\XFMaddressmarkstyle{\alph}

\begin{document}

\title{The {\sc Majorana} {\sc Demonstrator}: An R\&D project towards a tonne-scale germanium neutrinoless double-beta decay search}



\include{author_2009_06_MJ}

\begin{abstract}
The {\sc Majorana} collaboration is pursuing the development of
the so-called {\sc Majorana} {\sc Demonstrator}. The {\sc Demonstrator} is
intended to perform research and development towards a
tonne-scale germanium-based experiment to search for the
neutrinoless double-beta decay of $^{76}\mathrm{Ge}$. The
{\sc Demonstrator} can also perform a competitive direct dark matter
search for light WIMPs in the $1-10\,\mathrm{GeV}/c^2$ mass
range. It will consist of approximately 60~kg of germanium
detectors in an ultra-low background shield located deep
underground at the Sanford Underground
Laboratory in Lead, SD. The {\sc Demonstrator} will also perform background and technology studies,
and half of the  detector mass will be
enriched germanium. This talk will review the motivation, design,
technology and status of the Demonstrator.
\end{abstract}

\maketitle


\section{Introduction}

The \MJ\ collaboration intends to search for the neutrinoless double-beta decay (\BBz -decay) of the $^{76}\mathrm{Ge}$ isotope using High-purity germanium (HPGe) detectors. \BBz -decay is a currently unobserved\footnote{There exists a highly controversial claim of a discovery~\cite{kla04a, kla04b}. See~\cite{avi08} and references therein for a discussion of this claim.} nuclear decay where two neutrons in an atomic nucleus convert into two protons and two electrons with no neutrinos emitted. The existence of this process implies that neutrinos are Majorana fermions, ie. their own anti-particles~\cite{sche82}. It would also be the first observation of absolute lepton number violation. A related second-order weak process is two neutrino double-beta decay (\BBt -decay) where two antineutrinos are emitted as well. It constitutes an important background for \BBz\ searches. If we assume \BBz -decay is mediated by the exchange of a virtual, massive Majorana neutrino, the measured half-life of the decay will provide a measurement of the absolute neutrino mass scale, as opposed to the neutrino mass-squared differences from oscillation experiments. However, the conversion from a measured half-life to neutrino mass requires nuclear matrix elements that are difficult to calculate. 

An exciting new physics opportunity arose with the demonstration of the feasibility of HPGe p-type point contact (PPC) detector technology as dark matter detectors by the CoGeNT collaboration~\cite{Aal08}. PPC detectors have significantly reduced energy thresholds, from about \mbox{$5\,\mathrm{keV}$} to fractions of a keV, allowing a search for light \mbox{($<10\,\mathrm{GeV/c^2}$)} WIMP dark matter candidates that are inaccessible to current experiments. 

\section{Experimental Considerations}

Experiments that search for \BBz -decay face many challenges. The next generation of experiments will probe extremely rare decay rates, with half-lives on the order of $10^{26}$ to $10^{27}$ years.  These require a significant reduction in ionizing radiation backgrounds, necessitating deep underground sites, special materials selection and handling, mitigation of the high cost of enriching isotopes, and advanced analysis techniques. The goal is to achieve an extremely low, preferably near-zero background rate in the \BBz\ region-of-interest (ROI). The ROI is the range in the energy spectrum of the emitted electrons centered on the $Q$-value of the \BBz -decay, and a region around it determined by detector energy resolution. A specific goal of \MJ\ {\sc Demonstrator} is to demonstrate a background rate of 1 count per tonne-year exposure in a $4\,\mathrm{keV}$ wide ROI around the \gess\ $Q$-value of $2039\,\mathrm{keV}$ after all analysis cuts have been applied.

After moving the detector underground to remove backgrounds from cosmic-rays, the dominant remaining sources of backgrounds are from gamma-rays, surface alpha decays and beta-decays. A large fraction of these originate from natural isotope chains, primarily $^{40}\mathrm{K}$, $^{232}\mathrm{Th}$ and $^{238}\mathrm{U}$. Cosmic rays can activate certain materials at the surface prior to moving them underground, in particular creating radioactive $^{60}\mathrm{Co}$ in copper parts and $^{68}\mathrm{Ge}$ and $^{60}\mathrm{Co}$ in germanium. Hard neutrons from residual cosmic rays in the rock and shield can punch through the veto and produce counts in the detectors. Neutrons produced from $(\alpha,\mathrm{n})$ reactions in the rock are a background for dark matter searches. Monte Carlo estimates of acceptable levels of contaminants have played an important role in driving the design of the \MJ\ {\sc Demonstrator}. 

HPGe detectors constructed with germanium enriched in \gess\ are excellent candidates for the search of \BBz -decay. Germanium-based detectors have the best energy resolution of all \BBz -decay search technologies, which provides them with significant advantages. The high resolution improves the signal-to-background ratio, since a smaller ROI can be considered. Energy resolution also provides the only feasible discrimination against \BBt -decay events. Germanium detectors are intrinsically extremely clean and no special purification is required for \BBz -decay searches. Commercial technologies are well-established for the manufacturing of these detectors, and several options and vendors exist. 

\section{The \MJ\ {\sc Demonstrator}}

The \MJ\ experiment is a proposed, collaborative R\&D project involving nineteen institutions in four countries and about 50 scientists. \MJ\ requires a scaling of a factor of $\sim 100$ over what has been demonstrated in terms of background reduction and mass. Such a large improvement will require an intermediate step, the \MJ\  {\sc Demonstrator}. The initial module of the {\sc Demonstrator} will consist of 30~kg of 28~point-contact unenriched HPGe crystals mounted in an ultra-low background electroformed copper cryostat housed inside a massive, layered shield. The shield will, from the inside-out, consist of an electroformed inner copper shield, an OFHC copper shield, a lead shield, neutron moderator and a muon veto. The entire shield will also be enclosed in a radon-reduced volume. All materials are carefully selected to reduce radioactive backgrounds to acceptable levels. Its goal is to demonstrate background levels low enough to justify building a tonne-scale Ge experiment. Later, a 30~kg  enriched module will be installed test the recent claim of an observation of \BBz\ decay~\cite{kla04a, kla04b} and perform additional background measurements.

The initial \MJ\ modules will concentrate on P-PC Detectors that have the advantages of cost and simplicity, with no loss of physics reach.  Research and development of n-type segmented detectors and others will continue in parallel as an alternative. Several prototypes are under study by the collaboration and 18~additional natural germanium BeGe detectors have been ordered for the initial module. The {\sc Demonstrator} will be located at the 4850' level in the Sanford Underground Laboratory in Lead, South Dakota. 
The collaboration will also use the {\sc Demonstrator} to pursue longer term R\&D to minimize costs and optimize the schedule for a tonne-scale experiment. 


The infrastructure developed for the {\sc Demonstrator} at the Sanford Underground Laboratory will test several concept for the much more stringent and ambitious requirements for a tonne-scale experiments. The \MJ\ collaboration is developing, in collaboration with the Europen GERDA collaboration~\cite{sch05}, a suite of analysis and simulation tools. These tools will be tested with the {\sc Demonstrator} and its associated prototypes, and once tested, they will be valuable in the design and development of the tonne-scale detector. \MJ\ is also pursuing domestic capabilities of enrichment, one of the dominant cost drivers of a tonne-scale experiment. Finally, the performance of the GERDA experiment's liquid argon shielding technology and the \MJ\ experiment's traditional lead and copper shielding will be compared and a technology selected that optimizes the shield performance of the tonne-scale experiment. The GERDA and \MJ\ collaboration fully intend to merge to pursue the tonne-scale experiment. 
The anticipated half-life sensitivity of a tonne-scale experiment will be on the order of \mbox{$10^{28}\,\mathrm{years}$} depending on the background levels, which would completely probe the inverted neutrino mass-hierarchy region.


\begin{theacknowledgments}
The author gratefully acknowledges the support of DOE-NP under grant \# DE-FG02-97ER41041, the NSF under grant \# PHY-0705014, and the State of North Carolina.
\end{theacknowledgments}



\bibliographystyle{aipproc}   

\bibliography{main}

\IfFileExists{\jobname.bbl}{}
 {\typeout{}
  \typeout{******************************************}
  \typeout{** Please run "bibtex \jobname" to optain}
  \typeout{** the bibliography and then re-run LaTeX}
  \typeout{** twice to fix the references!}
  \typeout{******************************************}
  \typeout{}
 }

\end{document}

%% file: author_2009_06_MJ.tex
%
\newcommand{\alberta}{Centre for Particle Physics, University of Alberta, Edmonton, AB, Canada}
\newcommand{\blhill}{Department of Physics, Black Hills State University, Spearfish, SD, USA}
\newcommand{\ITEP}{Institute for Theoretical and Experimental Physics, Moscow, Russia}
\newcommand{\JINR}{Joint Institute for Nuclear Research, Dubna, Russia}
\newcommand{\lbnl}{Lawrence Berkeley National Laboratory, Berkeley, CA, USA}
\newcommand{\lanl}{Los Alamos National Laboratory, Los Alamos, NM, USA}
\newcommand{\queens}{Department of Physics, Queen's University, Kingston, ON, Canada}
\newcommand{\uw}{Center for Experimental Nuclear Physics and Astrophysics, and Department of Physics, University of Washington, Seattle, WA, USA}
\newcommand{\uchic}{Department of Physics, University of Chicago, Chicago, IL, USA}
\newcommand{\unc}{Department of Physics, University of North Carolina, Chapel Hill, NC, USA}
\newcommand{\ucne}{Department of Nuclear Engineering, University of California, Berkeley, CA, USA}
\newcommand{\ucph}{Department of Physics, University of California, Berkeley, CA, USA}
\newcommand{\duke}{Department of Physics, Duke University, Durham, NC, USA}
\newcommand{\ncsu}{Department of Physics, North Carolina State University, Raleigh, NC, USA}
\newcommand{\ornl}{Oak Ridge National Laboratory, Oak Ridge, TN, USA}
\newcommand{\osaka}{Research Center for Nuclear Physics and Department of Physics, Osaka University, Ibaraki, Osaka, Japan}
\newcommand{\pnnl}{Pacific Northwest National Laboratory, Richland, WA, USA}
\newcommand{\usc}{Department of Physics and Astronomy, University of South Carolina, Columbia, SC, USA}
\newcommand{\usd}{Department of Earth Science and Physics, University of South Dakota, Vermillion, SD, USA}
\newcommand{\utenn}{Department of Physics and Astronomy, University of Tennessee, Knoxville, TN, USA}
\newcommand{\tunl}{Triangle Universities Nuclear Laboratory, Durham, NC, USA}
\newcommand{\tunlNCSU}{Triangle Universities Nuclear Laboratory, Durham, NC, USA and Department of Physics, North Carolina State University, Raleigh, NC, USA}

\author{C.E.~Aalseth}{address={\pnnl}}
\author {M.~Amman}{address={\lbnl}}
\author{J.F.~Amsbaugh}{address={\uw}}
\author{F.T.~Avignone~III}{address={\usc}, altaddress= {\ornl}}
\author{H.O.~Back}{address={\ncsu} , altaddress= {\tunl}}
\author{A.S.~Barabash}{address={\ITEP}}
\author{P.S.~Barbeau}{address={\uchic}}
\author{J.R.~Beene}{address={\ornl}}
\author{M.~Bergevin}{address={\lbnl}}
\author{F.E.~Bertrand}{address={\ornl}}
\author{M.~Boswell}{address={\lanl}}
\author{V.~Brudanin}{address={\JINR}}
\author{W.~Bugg}{address={\utenn}}
\author{T.H.~Burritt}{address={\uw}}
\author{Y-D.~Chan}{address={\lbnl}}
\author{J.I.~Collar}{address={\uchic}}
\author{R.J.~Cooper}{address={\ornl}}
\author{R.~Creswick}{address={\usc}}
\author{J.A.~Detwiler}{address={\lbnl}}
\author{P.J.~Doe}{address={\uw}}
\author{Yu.~Efremenko}{address={\utenn}}
\author{V.~Egorov}{address={\JINR}}
\author{H.~Ejiri}{address={\osaka}}
\author{S.R.~Elliott}{address={\lanl}}
\author{J.~Ely}{address={\pnnl}}
\author{J.~Esterline}{address={\duke},altaddress= {\tunl}}
\author{H.~Farach}{address={\usc}}
\author{J.E.~Fast}{address={\pnnl}}
\author{N.~Fields}{address={\uchic}} 
\author{P.~Finnerty}{address={\unc}, altaddress= {\tunl}}
\author{B.~Fujikawa}{address={\lbnl}}
\author{E.~Fuller}{address={\pnnl}} 
\author{V.M.~Gehman}{address={\lanl}}
\author{G.K.~Giovanetti}{address={\unc}, altaddress={\tunl}}  
\author{V.E.~Guiseppe}{address={\lanl}}
\author{K.~Gusey}{address={\JINR}}
\author{A.L.~Hallin}{address={\alberta}}
\author{R.~Hazama}{address={\osaka}}
\author{R.~Henning \thanks{Corresponding Author}\ $^{,}$}{address={\unc}, altaddress= {\tunl}}
\author{A.~Hime}{address={\lanl}}
\author{E.W.~Hoppe}{address={\pnnl}}
\author{T.W.~Hossbach}{address={\usc}, altaddress= {\pnnl}}
\author{M.A.~Howe}{address={\unc}, altaddress= {\tunl}}
\author{R.A.~Johnson}{address={\uw}}
\author{K.J.~Keeter}{address={\blhill}}
\author{M.~Keillor}{address={\pnnl}}
\author{C.~Keller}{address={\usd}}
\author{J.D.~Kephart}{address={\tunlNCSU}, altaddress= {\pnnl}}
\author{M.F.~Kidd}{address={\duke}, altaddress= {\tunl}}
\author{O.~Kochetov}{address={\JINR}}
\author{S.I.~Konovalov}{address={\ITEP}}
\author{R.T.~Kouzes}{address={\pnnl}}
\author{K.T.~Lesko}{address={\lbnl}, altaddress= {\ucph}}
\author{L.~Leviner}{address={\ncsu}, altaddress= {\tunl}}
\author{J.C.~Loach}{address={\lbnl}}	
\author{P.N.~Luke}{address={\lbnl}}
\author{S.~MacMullin}{address={\unc} , altaddress= {\tunl}}
\author{M.G.~Marino}{address={\uw}}
\author{D.-M.~Mei}{address={\usd}}
\author{H.S.~Miley}{address={\pnnl}}
\author{M.~Miller}{address={\uw}} 
\author{L.~Mizouni}{address={\usc} , altaddress= {\pnnl}}  
\author{A.~Montoya}{address={\lanl}}  
\author{A.W.~Myers}{address={\uw}}
\author{M.~Nomachi}{address={\osaka}}
\author{B.~Odom}{address={\uchic}}
\author{J.L.~Orrell}{address={\pnnl}}
\author{D.~Phillips}{address={\unc}, altaddress= {\tunl}}  
\author{A.W.P.~Poon}{address={\lbnl}}
\author{G.~Prior}{address={\lbnl}}
\author{J. ~Qian}{address={\lbnl} } 
\author{D.C.~Radford}{address={\ornl}}
\author{K.~Rielage}{address={\lanl}}
\author{R.G.H.~Robertson}{address={\uw}}
\author{L.~Rodriguez}{address={\lanl}}
\author{K.P.~Rykaczewski}{address={\ornl}}
\author{A.G.~Schubert}{address={\uw}}
\author{T.~Shima}{address={\osaka}}
\author{M.~Shirchenko}{address={\JINR}}
\author{J.~Strain}{address={\unc} ,altaddress={\tunl}}
\author{K.~Thomas}{address={\usd}}		
\author{R.~Thompson}{address={\pnnl}}
\author{V.~Timkin}{address={\JINR}}
\author{W.~Tornow}{address={\duke} ,altaddress= {\tunl}}
\author{T. D.~Van Wechel}{address={\uw}}
\author{I.~Vanyushin}{address={\ITEP}}
\author{K.~Vetter}{address={\ucne} ,altaddress= {\lbnl}}
\author{R.~Warner}{address={\pnnl}}
\author{J.F.~Wilkerson}{address={\unc} , altaddress= {\tunl}}  
\author{J.M.~Wouters}{address={\lanl}}
\author{E.~Yakushev}{address={\JINR}}
\author{A.R.~Young}{address={\ncsu} , altaddress= {\tunl}}
\author{C.-H.~Yu}{address={\ornl}}
\author{V.~Yumatov}{address={\ITEP}}
\author{C.~Zhang}{address={\usd}}					
\author{S.~Zimmerman}{address={\lbnl}}  